\documentclass[10pt,aps,pra,twocolumn,superscriptaddress]{revtex4-2}

\usepackage{amsmath,amssymb,graphicx,bm,xcolor,hyperref}
\usepackage{url}
\begin{document}

\title{Inhomogeneous Synthetic Fluxes in Optical Lattices}

\author{Soujanya Datta}
\affiliation{Saha Institute of Nuclear Physics, 1/AF Bidhannagar, Kolkata 700064, India}
\affiliation{Homi Bhabha National Institute, Training School Complex, Anushakti Nagar, Mumbai 400094, India}
\author{Kush Saha}
\affiliation{National Institute of Science Education and Research, Jatni, 752050, India}
\affiliation{Homi Bhabha National Institute, Training School Complex, Anushakti Nagar, Mumbai 400094, India}
\author{Krishanu Roychowdhury}
\affiliation{Saha Institute of Nuclear Physics, 1/AF Bidhannagar, Kolkata 700064, India}
\affiliation{Homi Bhabha National Institute, Training School Complex, Anushakti Nagar, Mumbai 400094, India}

\date{\today}

\begin{abstract}
Synthetic gauge fields in optical lattices are usually engineered to realize a uniform magnetic flux. Here, we propose a scheme to generate spatially inhomogeneous fluxes and demonstrate how such fluxes reshape the single-particle spectrum. Within Wannier--Stark-assisted tunneling, onsite modulation produces a plaquette flux that depends on the orientation of the static tilt, whereas laser-assisted tunneling (LAT) does not. Using LAT, two coherent Raman processes with different momentum transfers imprint a dimerized Peierls phase that alternates the flux between neighboring plaquettes of a square lattice while keeping the mean flux fixed. A bond-selective Raman phase extends the scheme to the kagome lattice, where a parameter $\xi$ redistributes the flux between triangles and hexagons at fixed flux per unit cell. The Hofstadter--Rammal spectrum of the bipartite square lattice retains chiral symmetry for all $\xi$, whereas the kagome lattice lacks this symmetry, and its flux periodicity and reflection symmetries survive only at special values of $\xi$. As $\xi$ is reduced, the anomalous Landau levels of the kagome singular flat band are compressed toward the flat-band energy and evolve into a degenerate zero-energy manifold at $\xi=0$ with compactly localized states. The dynamical structure factor, accessible through Bragg scattering, resolves this evolution as a Rydberg-like series of peaks whose spectral span tracks the width of the anomalous-level tower and vanishes as $\xi\to0$. We discuss the frequency resolution and temperature required for its observation.

\end{abstract}

\maketitle

\section{Introduction}

Synthetic gauge fields in optical lattices \cite{Jaksch2003,bloch2008many,aidelsburger2011experimental, Aidelsburger2013, Miyake2013,goldman2016topological,moller2018synthetic} provide a controlled realization of lattice Hamiltonians with complex tunneling amplitudes, including the class with broken time-reversal symmetry. In analogy with a charged particle coupled to a vector potential, the gauge field is encoded in the tight-binding description through Peierls phases associated with tunneling between neighboring sites \cite{peierls1933theorie} -- the gauge-invariant phase accumulated around a closed plaquette is proportional to the magnetic flux through that plaquette. A uniform flux on a two-dimensional (2D) lattice realizes the Harper--Hofstadter (HH) Hamiltonian~\cite{harper1955single, Hofstadter1976}, a lattice realization of quantum-Hall physics characterized by the fractal Hofstadter--Rammal (HR) spectrum \cite{Hofstadter1976, rammal1985landau}. Such Peierls phases can be engineered using periodic driving and laser-assisted tunneling (LAT), providing direct control over the effective magnetic flux through the engineered tunneling processes \cite{Jaksch2003,aidelsburger2011experimental, Miyake2013, Aidelsburger2013,goldman2016topological}.

While most implementations aim at generating a uniform flux \cite{Jaksch2003,aidelsburger2011experimental, Aidelsburger2013, Miyake2013,goldman2016topological}, inhomogeneous fluxes~\cite{shi1997energy, iye2004hofstadter, wang2007redistributing} offer an additional degree of control. Spatial modulation of the flux can reconstruct magnetic Bloch bands (MBBs), alter their topology, and, for suitable patterns, generate flat bands or localization through destructive interference \cite{Datta2025}. Flux inhomogeneity also affects the formation of Landau levels (LLs). In lattices whose zero-field band structure contains a singular flat band, a uniform field produces anomalous Landau levels (ALLs) below the flat-band energy, with an unconventional $E_n\propto -1/n$ dependence \cite{Rhim2020}; a spatially textured flux modifies this tower \cite{Datta2025}. These possibilities motivate protocols that control both the magnitude and spatial pattern of the plaquette flux independently of the geometric constraints imposed by the driving scheme.

We therefore ask how inhomogeneous fluxes can be enabled in optical lattices. One route is Wannier--Stark-assisted tunneling \cite{Gluck2002,sias2008observation}, where a static tilt creates energy offsets between neighboring sites and periodic driving restores resonant tunneling while imprinting phases on the hopping. In an optical lattice, this can be implemented either through onsite-energy modulation, thereby inducing photon-assisted tunneling, or by LAT, in which laser fields resonantly couple neighboring sites and generate spatially dependent hopping phases \cite{Aidelsburger2013, Miyake2013}. We illustrate both approaches on the square lattice for a uniform static tilt of arbitrary orientation. With onsite modulation, the resulting complex hoppings produce a uniform synthetic flux whose magnitude depends on the tilt orientation, whereas for LAT, the flux is insensitive to it.

We next introduce spatially inhomogeneous fluxes, first on the square lattice, the standard setting for Hofstadter physics \cite{Hofstadter1976} in optical lattices, that can be generated by two orthogonal pairs of counter-propagating beams \cite{hemmerich1991light,hemmerich1993two,bloch2005ultracold,lim2008staggered}, producing the separable potential $V(x,y)=V_x\sin^2(2\pi x/\lambda)+V_y\sin^2(2\pi y/\lambda)$ with lattice spacing $a=\lambda/2$. We consider a dimerized pattern in which the plaquette flux alternates along $x$ at fixed mean flux, in the spirit of the modulated and staggered fluxes studied on the square lattice in condensed matter~\cite{shi1997energy, iye2004hofstadter, wang2007redistributing} and realized with cold atoms~\cite{lim2008staggered, aidelsburger2011experimental}; it leads to an enlarged magnetic unit cell and a reconstruction of the HR spectrum. We then turn to the kagome lattice, where the interplay between the inhomogeneous flux pattern and the multiband structure, including a flat band, gives rise to a qualitatively different LL spectrum. In particular, we study ALLs, which populate the energy window where the zero-field band structure has no dispersion, and show that the flux inhomogeneity controls their spectral width, including their collapse onto a degenerate zero-energy manifold for suitable flux distributions. As shown in Ref. \cite{Datta2025}, this collapse is rooted in an emergent lattice supersymmetry (SUSY); here we concentrate on its signatures in the spectrum and in the response.

Finally, we address the spectroscopic detection of the ALLs using Bragg scattering \cite{weidemuller1995bragg,rey2005bragg,ernst2010probing}. The Bragg response, determined by the dynamical structure factor (DSF), provides momentum- and frequency-resolved information about transitions between the MBBs, and resolves the evolution of the ALLs with the flux modulation.

The paper is organized as follows. In Sec. \ref{sec:ws1d}, we review Wannier--Stark physics in one dimension (1D). In Sec. \ref{sec:onsite1d}, we use it to generate a synthetic gauge field by onsite modulation. In Secs. \ref{sec:ws2d} and \ref{sec:onsite2d}, we extend the construction to a tilted square lattice and show that the resulting flux depends on the orientation of the tilt, in contrast to LAT. In Sec. \ref{sec:sqlat}, we introduce a dimerized inhomogeneous flux on the square lattice through two coherent LAT processes. In Sec. \ref{sec:kaglat}, we extend the scheme to the kagome lattice. In Sec. \ref{sec:hr}, we analyze the HR spectra on both lattices, with emphasis on the ALLs. In Sec. \ref{sec:dsf}, we study their detection through the DSF, before concluding in Sec. \ref{sec:concl}.

\section{Wannier--Stark scaffold in 1D}\label{sec:ws1d}

We begin by reviewing the Wannier--Stark physics of a 1D lattice for completeness. The tight-binding Hamiltonian, with uniform nearest-neighbor (nn) hopping amplitude $t$ and exposed to a static electric field $\mathcal E$, i.e., a uniform tilt of the lattice potential, is
\begin{align}\label{ham1D}
    \mathcal H_{\rm 1D} = -t \sum_l \left[|l\rangle\langle l+1| + {\rm h.c.}\right] + \mathcal{E}a\sum_l l\,|l\rangle\langle l|\,,
\end{align}
where $a$ is the lattice spacing and $|l\rangle$ the single-particle state at site $l$. An eigenstate $|\psi\rangle=\sum_l\phi_l|l\rangle$ at energy $E$ satisfies
\begin{align}\label{recusrln1}
 -t\left(\phi_{l+1} + \phi_{l-1} \right) + \left(l\mathcal{E}a-E\right)\phi_l=0.
\end{align}
With the translation operator $T|l\rangle=|l+1\rangle$, one has $T^{-1}\mathcal H_{\rm 1D}T=\mathcal H_{\rm 1D}+\mathcal Ea$. Hence the Schr\"{o}dinger equation $\mathcal H_{\rm 1D}|\psi\rangle=E|\psi\rangle$ translates to $\mathcal H_{\rm 1D}\,T|\psi\rangle=(E+\mathcal Ea)\,T|\psi\rangle$, and the spectrum is a ladder, $E(n)=n\mathcal Ea$, with integer $n$. Inserting this in Eq.~\eqref{recusrln1} and defining $d=l-n$ and $\varphi_d\equiv\phi_l$, we obtain the solution in terms of the Bessel functions, $\varphi_d=J_d(2t/\mathcal{E}a)$. The $n$-th eigenstate is $|\psi_n\rangle=\sum_l J_{l-n}(2t/\mathcal{E}a)|l\rangle$, localized around $l=n$.

In the continuum limit $a\to0$, with $x=la$, $\phi_l\equiv\phi(x)$ and $\phi_{l\pm1}=\phi\pm a\phi'+(a^2/2)\phi''+\mathcal O(a^3)$, Eq.~\eqref{recusrln1} becomes
\begin{align}
 -ta^2 \phi{''}(x) + \mathcal{E}x\, \phi(x) = (E+2t)\, \phi(x),
\end{align}
solved by the Airy function $\phi(x)={\rm Ai}[(x-x_0)/\ell_W]$ with $x_0=(E+2t)/\mathcal E$ and $\ell_W=(ta^2/\mathcal E)^{1/3}$. The discrete ladder is then replaced by a continuous spectrum.

\section{Synthetic gauge field from onsite modulation in 1D}\label{sec:onsite1d}

The Bessel profile of Sec.~\ref{sec:ws1d} shows that each Wannier--Stark state extends over only $\sim2t/\Delta$ sites, where $\Delta=\mathcal Ea$ is the energy offset between neighboring sites (referred to as the \emph{Stark offset}); for $\Delta\gg t$, the tunneling is effectively frozen. This localization is the starting point for engineering gauge fields: a periodic drive at frequency $\omega=\Delta$ supplies the missing energy, restores tunneling between adjacent sites through photon-assisted processes, and imprints a controllable phase on each tunneling amplitude. Such a drive may enter either through the onsite energies or directly through LAT, providing two routes to engineer the Peierls phases.

For LAT, a pair of Raman lasers with frequency difference $\omega=\omega_1-\omega_2$, resonant with the Stark offset, induces tunneling between adjacent sites. The Raman process transfers a momentum $\tilde{\mathbf q}=\mathbf k_1-\mathbf k_2$, so the coupling carries a phase $(\tilde{\mathbf q}\cdot\mathbf x-\omega t+\phi_0)$, assigned to the effective hopping.

Other optical techniques modulate the onsite energies directly, for example by superimposing an optical potential whose spatial profile is translated in time, realized with a focused laser, a digital micromirror device (DMD), or an optical superlattice with a time-dependent phase offset. We model such a perturbation by $V_l(t)=V_0\cos(\omega t-\tilde q la+\phi_0)$, so that
\begin{align}
   \tilde{\mathcal H}_{\rm 1D} = \mathcal H_{\rm 1D} + \sum_l V_l(t)\,|l\rangle\langle l|,
\end{align}
with $\omega$ adjusted to the resonance $\omega=\Delta$ (more generally $\Delta=j\omega$, see below).

We transform to a rotating frame, $|\psi\rangle=U(t)|\tilde\psi\rangle$, using two unitaries, $U_1(t)=\exp[-i\Delta t\sum_l l|l\rangle\langle l|]$ and $U_2(t)=\exp[-i\sum_l\chi_l(t)|l\rangle\langle l|]$ with $\dot\chi_l=V_l$. The first removes the tilt $\Delta$ and the second the modulation $V_l(t)$, transferring both into time-dependent phases of the hopping amplitudes:
\begin{align}\label{ham1Db}
    \tilde{\mathcal H}^{(2)}_{\rm 1D} = -t \sum_l \Big[ e^{-i[\Delta t+\chi_{l+1}(t)-\chi_l(t)]} |l\rangle\langle l+1| + {\rm h.c.}\Big],
\end{align}
with $\chi_{l+1}(t)-\chi_l(t)$
\[
= -2(V_0/\omega)\sin(\tilde{q} a/2)\cos\left(\omega t - \tilde{q}la + \phi_0 - \tilde{q}a/2 \right).
\]
Using the Jacobi--Anger expansion, $e^{iz\cos\vartheta}=\sum_j i^jJ_j(z)e^{ij\vartheta}$ with $z=(2V_0/\omega)\sin(\tilde qa/2)$, the hopping factor becomes $\sum_{j}i^jJ_j(z)\,e^{-ij[\tilde q(l+1/2)a-\phi_0]} e^{i(j\omega-\Delta)t}$. At the $j$-photon resonance $\Delta=j\omega$, the $j$-th term is static, while all others oscillate at multiples of $\omega$ and are dropped in the rotating-wave approximation \cite{goldman2014periodically,eckardt2017colloquium}. The effective Hamiltonian is
\begin{align}
\tilde{\mathcal H}^{(2)}_{\rm 1D}
={}&-t\sum_l\Big[ i^{j}J_j(z)\,e^{-ij[\tilde q(l+1/2)a-\phi_0]} |l\rangle\langle l+1|+{\rm h.c.}\Big].
\end{align}
Restricting to the one-photon resonance, $\Delta=\omega$ ($j=1$), and retaining the leading contribution in the weak-driving regime $V_0/\omega\ll1$ (in which 
$J_j(z)\sim (z/2)^j/j!$ drops with $j$) gives
\begin{align}
\tilde{\mathcal H}^{(2)}_{\rm 1D}
={}&-t\,J_1\!\left(\frac{2V_0}{\omega}\sin\frac{\tilde{q}a}{2}\right) \nonumber \\
& \times \sum_l
\Big[\,i\,e^{-i[\tilde{q}(l+1/2)a-\phi_0]}\,|l\rangle\langle l+1|+\mathrm{h.c.}\Big].
\end{align}
Barring the spatially uniform phase $(\pi/2-\tilde{q}a/2+\phi_0)$, which can be absorbed by a gauge transformation, the resulting tunneling carries the spatially varying Peierls phase \(\phi_l=-\tilde{q}la\), which can be identified as a synthetic gauge field in 1D. Although no gauge-invariant magnetic flux can be defined in 1D, such bond-dependent phases provide the building blocks for synthetic magnetic fluxes in higher dimensions, which become manifest in 2D optical lattices, where closed plaquettes enclose a gauge-invariant flux.

\section{An excursion to 2D: Wannier-Stark physics on a square lattice}\label{sec:ws2d}

We now extend the construction to a square lattice with sites at $\mathbf x=a(m,n)$ and the field ${\cal E}$ along a commensurate direction specified by two coprimes $(q_E,p_E)$. For such rational directions, the Wannier--Stark eigenstates are known to be localized along the field and extended normal to it~\cite{gluck2001wannier, kolovsky2003bloch}; for the separable nn lattice considered here, we derive them in closed form, which also yields the bond offsets used in Sec.~\ref{sec:onsite2d}. The tight-binding Hamiltonian is
\begin{align}\label{ham2Da}
  \mathcal H_{\rm 2D} = & -t \sum_{m,n} \Big[|m,n\rangle\langle m+1,n| + |m,n\rangle\langle m,n+1| + {\rm h.c.}\Big] \nonumber\\
  &+ \sum_{m,n} V_{m,n}|m,n \rangle\langle m,n|\,,
\end{align}
with $V_{m,n}=\Delta(m\cos\vartheta+n\sin\vartheta)$, $\Delta=\mathcal Ea$, and $\tan\vartheta=p_E/q_E$. In the coordinates $u=q_Em+p_En$ and $v=-p_Em+q_En$ the tilt is $V=F_uu$, independent of $v$, with $F_u=\Delta\cos\vartheta/q_E=\Delta/\sqrt{p_E^2+q_E^2}$. A hop $(m,n)\to(m+1,n)$ changes $(u,v)\to(u+q_E,v-p_E)$, and a hop $(m,n)\to(m,n+1)$ changes $(u,v)\to(u+p_E,v+q_E)$.

Since the onsite potential depends only on $u$, we expand $|\psi\rangle=\sum_{u,v}e^{ik_vv}\phi_u|u,v\rangle$ (taking Bloch waves in the transverse direction $v$), and obtain the recurrence
\begin{align}\label{recrln2Da}
 &-t\Big[ e^{-ik_vp_E}\phi_{u+q_E} + e^{ik_vp_E}\phi_{u-q_E} \nonumber\\
 &\hphantom{-t\Big[}+ e^{ik_vq_E}\phi_{u+p_E} + e^{-ik_vq_E}\phi_{u-p_E} \Big] \nonumber\\
 &+ (F_uu-E)\phi_u = 0\,.
\end{align}
This is a generalized 1D Wannier--Stark problem with hoppings over distances $\pm p_E$ and $\pm q_E$. Because $p_E$ and $q_E$ are coprime, there is a lattice translation that shifts $u$ by one unit and shifts the tilt by $F_u$, so the argument of Sec.~\ref{sec:ws1d} gives the ladder $E(l)=lF_u$ with integer $l$. Defining $\kappa_q=k_vp_E$ and $\kappa_p=-k_vq_E$, $h=u-l$ and $\varphi_h\equiv\phi_{h+l}$, we obtain
\begin{align}\label{recrln2Db}
  &t\Big[ e^{-i\kappa_q}\varphi_{h+q_E} + e^{i\kappa_q}\varphi_{h-q_E} + e^{-i\kappa_p}\varphi_{h+p_E} + e^{i\kappa_p}\varphi_{h-p_E} \Big] \nonumber\\
  &= F_uh\,\varphi_{h}\,.
\end{align}
In terms of the Laurent series $G(z)=\sum_{h\in\mathbb Z}\varphi_hz^h$, this reads
\begin{align}\label{recrln2Dc}
  &t\Big[ e^{-i\kappa_q}z^{-q_E} + e^{i\kappa_q}z^{q_E} + e^{-i\kappa_p}z^{-p_E} + e^{i\kappa_p}z^{p_E} \Big]G \nonumber\\
  &= F_uz\,\frac{dG}{dz}\,.
\end{align}
Defining $w_q=e^{i\kappa_q}z^{q_E}$ and $w_p=e^{i\kappa_p}z^{p_E}$, Eq.~\eqref{recrln2Dc} integrates to
\begin{align}
    G(z) ={}& \exp\!\Big[\frac{t}{q_EF_u}\Big(w_q-\frac{1}{w_q}\Big)\Big] \exp\!\Big[\frac{t}{p_EF_u}\Big(w_p-\frac{1}{w_p}\Big)\Big].
\end{align}
Using $\exp[\tfrac x2(w-1/w)]=\sum_hJ_h(x)w^h$ and writing $J^{(q)}_h\equiv J_h(2t/q_EF_u)$ and $J^{(p)}_{h'}\equiv J_{h'}(2t/p_EF_u)$,
\begin{align}
    G(z) ={}& \sum_{h,h'} J^{(q)}_hJ^{(p)}_{h'}  e^{i(h p_E-h' q_E)k_v}z^{hq_E+h'p_E},
\end{align}
and therefore
\begin{align}
    \varphi_j ={}& \sum_{h,h'} J^{(q)}_hJ^{(p)}_{h'}  e^{i(h p_E-h' q_E)k_v}\,\delta_{j,\,hq_E+h'p_E}\,,
\end{align}
which has a localization content similar to the 1D problem. The ladder $E(l)=lF_u$ is independent of $k_v$, i.e., the band normal to the field is flat; this infinite degeneracy, specific to separable lattices, allows localized combinations in both directions as an alternative eigenbasis~\cite{kolovsky2003bloch}.

\section{Synthetic fluxes on the tilted square lattice from onsite modulation}\label{sec:onsite2d}

The 2D Wannier--Stark problem provides the starting point for generating synthetic fluxes by onsite modulation~\cite{gerbier2010gauge,moller2010condensed}, as in Sec.~\ref{sec:onsite1d}. We add to Eq.~\eqref{ham2Da} the term $\sum_{m,n}\tilde V_{m,n}(t)|m,n\rangle\langle m,n|$ with $\tilde V_{m,n}(t)=V_0\cos(\omega t-k_xma-k_yna)$. A uniform $\tilde V$ is a pure gauge and generates no flux.

The $x$ bonds have a Stark offset $\Delta_x=q_EF_u$ and the $y$ bonds $\Delta_y=p_EF_u$. Proceeding as in Sec.~\ref{sec:onsite1d}, with $\dot\chi_{m,n}=\tilde V_{m,n}$, the rotating-frame hoppings acquire the factors $e^{-i(\Delta_xt+\chi_{m+1,n}-\chi_{m,n})}$ along $x$ and $e^{-i(\Delta_yt+\chi_{m,n+1}-\chi_{m,n})}$ along $y$. The Jacobi--Anger expansion has resonant terms when $j_x\omega=q_EF_u$ and $j_y\omega=p_EF_u$, with integer orders $j_x,j_y$, so $j_x/j_y=q_E/p_E$. For coprime $(p_E,q_E)$, the lowest-order resonance is $j_x=q_E$, $j_y=p_E$ at $\omega=F_u$, and the effective hopping amplitudes are
\begin{align}
t_x = -t\,i^{j_x}J_{j_x}\!\Big(\frac{2V_0}{\omega}\sin\frac{k_xa}{2}\Big)~;~t_y = -t\,i^{j_y}J_{j_y}\!\Big(\frac{2V_0}{\omega}\sin\frac{k_ya}{2}\Big),
\end{align}
which are in general unequal, so the resulting HH model is anisotropic. The Peierls phases are $-j_x[k_x(m+\tfrac12)+k_yn]a$ on the $x$ bond leaving $(m,n)$ and $-j_y[k_xm+k_y(n+\tfrac12)]a$ on the $y$ bond leaving $(m,n)$. Summing them around a plaquette gives the phase $(F_ua/\omega)(q_Ek_y-p_Ek_x)$, i.e., the uniform flux
\begin{align}
\Phi_P = \frac{\Phi_0}{2\pi}\,\frac{F_ua}{\omega}\,(q_Ek_y-p_Ek_x),
\end{align}
whose magnitude depends on the orientation $(p_E,q_E)$ of the tilt, as in earlier photon-assisted-tunneling schemes~\cite{kolovsky2011creating, creffield2014generation}. (The overall sign is a convention, as in Sec.~\ref{sec:onsite1d}.) This orientation dependence comes at a practical cost: the effective hoppings involve Bessel functions of orders $j_x=q_E$ and $j_y=p_E$, which for weak driving ($z \sim V_0/\omega \ll 1$), scale as $(z/2)^{j_{x,y}}/j_{x,y}!$, so that only the lowest orientations, such as $(p_E,q_E)=(0,1)$ and $(1,1)$, retain sizable tunneling. This motivates the use of LAT for the inhomogeneous fluxes considered below.

For LAT, the Raman frequency differences are set equal to the bond offsets, $\omega_x=q_EF_u$ and $\omega_y=p_EF_u$, so that the Raman phase factors $e^{i\omega_{x,y}t}$ cancel the tilt phases. The phase imprinted on each bond is then $\tilde{\mathbf q}_{x,y}\cdot\mathbf x$, set by the Raman momentum transfers $\tilde{\mathbf q}_{x,y}$ on the $x$ and $y$ bonds alone. A uniform nonzero flux results, but it is independent of $(p_E,q_E)$ and therefore cannot be tuned through the orientation of the tilt.

\section{Inhomogeneous fluxes on the square lattice through LAT}\label{sec:sqlat}

\begin{figure}[t]
    \centering
    \includegraphics[width=\columnwidth]{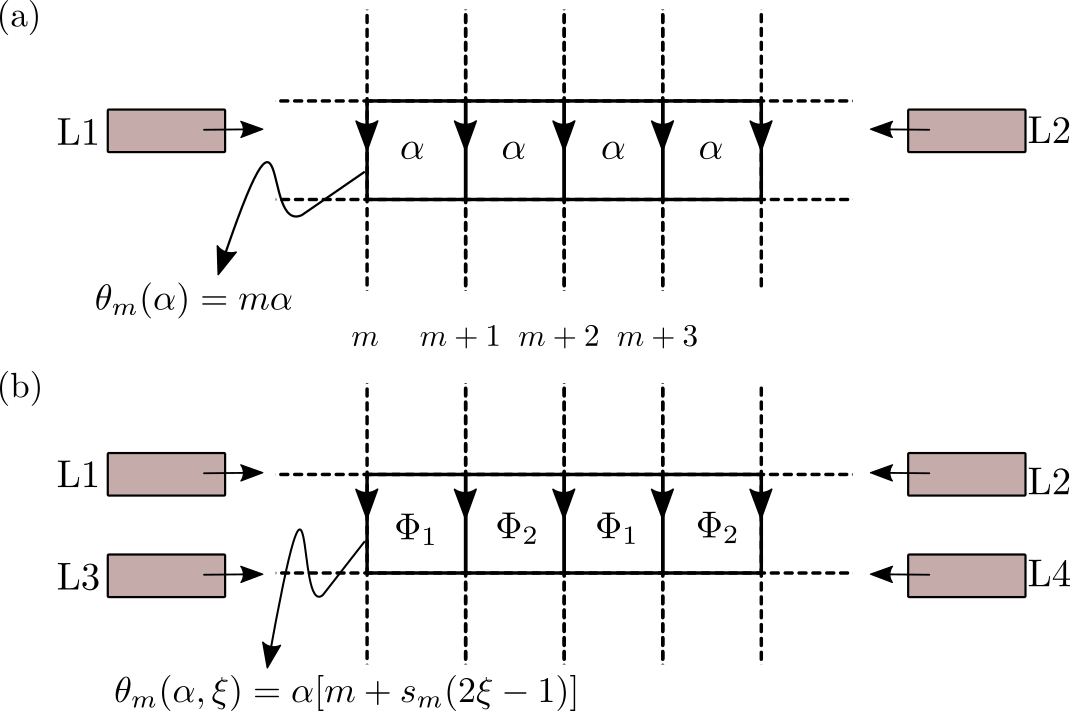}
    \caption{LAT on the vertical bonds of a square optical lattice, which are suppressed by a static tilt along $y$. The integer $m$ labels the columns, and the arrows on the vertical bonds indicate the direction of the Peierls phase $\theta_m$. (a) A Raman pair L1, L2 with momentum transfer $\tilde{\mathbf q}=\mathbf k_{\rm L1}-\mathbf k_{\rm L2}=(2\pi\alpha/\Phi_0a,0)$ imprints the phase $\theta_m(\alpha)=(\Phi_0/2\pi)\,\tilde{\mathbf q}\cdot\mathbf x_m=m\alpha$, with $\mathbf x_m=(ma,0)$, producing a uniform flux $\alpha$ per plaquette. (b) Two coherent Raman pairs, L1--L2 and L3--L4, with momentum transfers $\tilde{\mathbf q}_1$ and $\tilde{\mathbf q}_2$, contribute to the tunneling amplitude with complex weights $A$ and $B$ determined by $\xi$ and $\alpha$ [Eq.~\eqref{eq:two_laser_phase}]. The resulting Peierls phases, $\theta_m(\alpha,\xi)=\alpha[m+s_m(2\xi-1)]$ [Eq.~\eqref{dimerized_phase}], give alternating plaquette fluxes $\Phi_1=2\xi\alpha$ and $\Phi_2=2(1-\xi)\alpha$ for odd and even $m$ respectively [Eq.~\eqref{dimerized_flux2}]; the flux is homogeneous at $\xi=1/2$.}
    \label{fig1}
\end{figure}

The spatial structure of the LAT processes, however, provides a route to extending this construction beyond uniform fluxes. In particular, by allowing the momentum transfer and/or the phase of the Raman coupling to vary spatially, the Peierls phases acquired along different bonds can be engineered independently. The resulting plaquette flux is then spatially modulated, providing a direct route to generating inhomogeneous synthetic magnetic fluxes. We demonstrate this first on the square lattice, with a dimerized flux pattern, before extending it to the kagome lattice in Sec.~\ref{sec:kaglat}.

Here and below, we write Peierls phases in flux units: a bond labeled by $\theta$ carries the hopping factor $e^{2\pi i\theta/\Phi_0}$, so that the flux through a plaquette is the sum of the $\theta$'s around it.
The simplest realization of a uniform synthetic flux $\alpha$ per plaquette is obtained by engineering the Peierls phases in the Landau gauge, for which the hopping along one lattice direction is real (here we take $x$) while that along the orthogonal direction ($y$) acquires a position-dependent phase. In the LAT scheme, such a gauge can be realized by an appropriate choice of the momentum transfer between a pair of Raman beams L1 and L2, $\tilde{\bf q}={\bf k}_{\rm L1}-{\bf k}_{\rm L2}=(2\pi\alpha/\Phi_0a,0)$ [Fig.~\ref{fig1} (a)]. In particular, the component of $\tilde{\bf q}$ along the $x$ direction determines the spatial variation of the Peierls phase through the gauge field ${\bf A}=(0,-m\alpha)$ ($m$ denoting the $x$-coordinates of the sites as $x_m=ma$), and hence the flux accumulated around a plaquette. Choosing this component such that the Peierls phase on successive vertical bonds differs by $\alpha$ realizes a uniform flux $\alpha$ per plaquette and consequently the conventional Hofstadter Hamiltonian.

The same flexibility allows a continuous deformation into an inhomogeneous configuration. Rather than changing the average flux, one redistributes the flux between neighboring plaquettes by adding a staggered component to the $y$-bond phase:
\begin{align}\label{dimerized_phase}
\theta_m(\alpha,\xi) = \alpha\left[m+s_m(2\xi-1)\right],~s_m=\frac{1+(-1)^m}{2},
\end{align}
where $\xi$ controls the degree of modulation and $s_m=1\,(0)$ for even (odd) $m$. The plaquette flux is
\begin{align}\label{dimerized_flux}
\Phi^{\rm pl}_m(\alpha,\xi)&=\theta_{m+1}(\alpha,\xi)-\theta_m(\alpha,\xi) \nonumber \\  &=\alpha\left[1-(2\xi-1)(-1)^m\right],
\end{align}
that is,
\begin{equation}\label{dimerized_flux2}
\Phi^{\rm pl}_m(\alpha,\xi)=
\begin{cases}
2\alpha(1-\xi), & m\ {\rm even},\\
2\alpha\xi, & m\ {\rm odd}.
\end{cases}
\end{equation}
For $\xi=1/2$ the flux is homogeneous, $\Phi^{\rm pl}_m=\alpha$, while varying $\xi$ redistributes the flux without changing the average $(\Phi^{\rm pl}_{\rm even}+\Phi^{\rm pl}_{\rm odd})/2=\alpha$. The limits $\xi=0$ and $\xi=1$ correspond to alternating fluxes $2\alpha$ and $0$, with the order of the two plaquette types interchanged.

The required spatial modulation of the Peierls phase can be generated by extending the single Raman process used for the uniform Landau gauge to two coherent LAT processes with different momentum transfers [Fig.~\ref{fig1} (b)]. Defining $\beta=\alpha(2\xi-1)$, Eq.~\eqref{dimerized_phase} can be written as
\begin{align}
e^{2\pi i\theta_m/\Phi_0}
={}&e^{2\pi i\alpha m/\Phi_0}
\Big[
\frac{1+e^{2\pi i\beta/\Phi_0}}{2} \nonumber\\
&+\frac{e^{2\pi i\beta/\Phi_0}-1}{2}\,e^{i\pi x_m/a}
\Big],
\label{eq:two_laser_phase}
\end{align}
where the two terms correspond to Raman momentum transfers
\begin{equation}
\tilde{\bf q}_1=\left(\frac{2\pi\alpha}{\Phi_0a},0\right);~
\tilde{\bf q}_2={\bf k}_{\rm L3}-{\bf k}_{\rm L4}=\left(\frac{2\pi\alpha/\Phi_0+\pi}{a},0\right).
\label{eq:dimerized_momentum}
\end{equation}
Accordingly, when both Raman pairs are tuned to resonance (both frequencies match the energy offset), choosing the complex amplitudes of the two coherent Raman processes as $A=(1+e^{2\pi i\beta/\Phi_0})/2$, $B=(e^{2\pi i\beta/\Phi_0}-1)/2$, gives the required spatial dependence of the Peierls phase. Their magnitudes are $|A|=|\cos(\pi\beta/\Phi_0)|$, $|B|=|\sin(\pi\beta/\Phi_0)|$, while their relative optical phase is fixed by $\beta$. Thus, varying $\xi$ continuously redistributes the weight between the two momentum-transfer components, providing a controlled interpolation between the homogeneous and the dimerized inhomogeneous-flux configuration. In the homogeneous case, $B=0$, and the second Raman process is absent. 

\section{Inhomogeneous fluxes on a kagome optical lattice}\label{sec:kaglat}

\begin{figure*}[t]
\centering
\includegraphics[width=1.8\columnwidth]{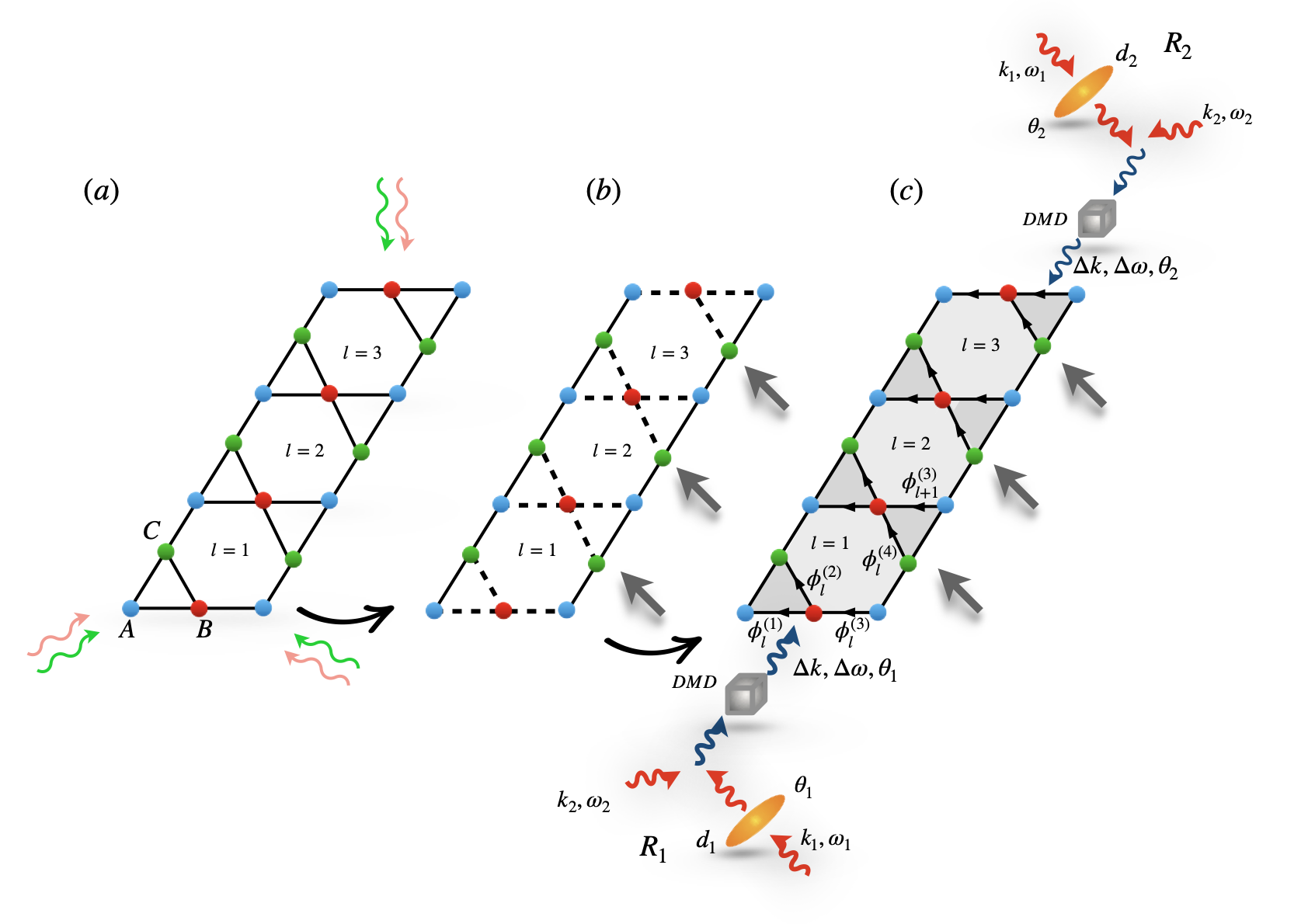}
    \caption{Schematic of the LAT protocol for generating a flux-inhomogeneous kagome optical lattice. The structural cells $l=1,2,3$ form the magnetic unit cell (here $q=3$), and the sites $A$, $B$, $C$ are shown in blue, red, and green.
    (a) Kagome lattice without a tilt, where the nn hoppings along the $AB$, $BC$, and $AC$ bonds are equivalent.
    (b) A static tilt (gray arrows) perpendicular to the $AC$ bonds shifts the energy of the $B$ sites relative to the $A$ and $C$ sites, suppressing tunneling along the $AB$ and $BC$ bonds (dashed lines), while the $AC$ bonds remain resonant.
    (c) The suppressed hoppings are restored by two Raman pairs, $R_1$ and $R_2$, with momentum and frequency differences $(\Delta\mathbf k,\Delta\omega)$. DMDs restrict $R_1$ to the lower and $R_2$ to the upper triangles. The optical phases $\theta_1$ and $\theta_2$ of the two pairs are the phases $\theta^{L}_{BC}$ and $\theta^{U}_{BC}$ imprinted on the $BC$ links of the lower and upper triangles, respectively. Arrows mark the bonds carrying the net phases $\phi^{L,U}_{AB}$ and $\phi^{L,U}_{BC}$ of Eq.~\eqref{bondphase1}, and the shading indicates the flux density in the triangular (dark) and hexagonal (light) plaquettes. Notation as in Sec.~\ref{sec:kaglat}.}
    \label{fig2}
\end{figure*}
The consequences of inhomogeneous fluxes are richer in a multiband setting, in particular in the presence of a gapless flat band, where the absence of dispersion and the topological band degeneracy provide a distinct platform for flux-induced spectral reconstruction. The kagome optical lattice is a natural setting for this purpose, which can be generated with superlattices that combine laser beams of two wavelengths \cite{Jo2012,hui2019flux}.

A static field of strength $\mathcal E$ along $\hat e_\perp=(\sqrt3/2,-1/2)$, perpendicular to the $AC$ bonds of the kagome lattice (Fig.~\ref{fig2}), tilts the lattice potential and shifts the onsite energies to $E_i=\mathcal E\,\hat e_\perp\cdot\mathbf X_i$, with $\mathbf X_i$ the position of site $i$. The $A$ and $C$ sites then remain degenerate, while each $B$ site is detuned from its $A$ and $C$ neighbors by $\Delta=\sqrt3\,\mathcal Ea/2$, lying above them in the upper triangles and below them in the lower ones. The tilt thus suppresses tunneling along the $AB$ and $BC$ bonds of all triangles, whereas the $AC$ bonds, transverse to the field, stay resonant. The suppressed processes are restored by LAT with two coherent Raman pairs, each addressing the $AB$ and $BC$ bonds of one set of triangles: the first pair is active on the lower triangles and suppressed on the upper ones, and conversely for the second.

\subsection{Peierls phases from LAT}

Since the $AC$ bonds are resonant and real, the flux through a triangle is determined by the phases of its $AB$ and $BC$ bonds. These share the Stark offset $\Delta$ and are therefore restored by the same Raman resonance (the Raman phase for a hop up the tilt is conjugate to that for the hop down the tilt). Since the Raman phase of a bond is evaluated at its center, the Raman contribution to the flux of a triangle is $(\Phi_0/4\pi)\,\Delta\mathbf k\cdot(\mathbf x_A-\mathbf x_C)$, up to the orientation sign. For $\Delta\mathbf k=(2\pi\alpha/\Phi_0a)(1,\sqrt3)$, which points along the $AC$ bond with $\mathbf x_A-\mathbf x_C=-\tfrac a2(1,\sqrt3)$, this contribution equals $\alpha$. Together with the corresponding $6\alpha$ through each hexagon, the Raman phases alone thus produce the homogeneous pattern, with $8\alpha$ per unit cell, independently of $\xi$; here and below, fluxes are noted up to a common orientation sign.

To redistribute this flux at fixed $8\alpha$ per unit cell, we add optical phases to the bonds and parametrize the bond phase as $\phi^{U/L}_{\rm bond}=(\Phi_0/2\pi)\,\Delta\mathbf k\cdot\mathbf x+\theta^{U/L}_{\rm bond}$ for the upper (U) and lower (L) triangles of a unit cell, with
\begin{align}\label{bondphase1}
 \theta_{AB}^{U/L}=0~;~\theta_{BC}^{U}=-\theta_{BC}^{L}=-\alpha(1-\xi),
\end{align}
which are the optical phases $\theta_{2}=\theta_{BC}^{U}$ and $\theta_1=\theta_{BC}^{L}$ of Fig.~\ref{fig2}. With the phase carried by the hop up the tilt, $\theta_{BC}$ enters each triangle with the sign opposite to the Raman contribution: it subtracts $(1-\xi)\alpha$ from each triangle and adds $2(1-\xi)\alpha$ to each hexagon, so that
\begin{align}\label{fluxkag}
 \Phi_\triangle(\alpha,\xi)=\xi\alpha,\qquad \Phi_{\rm hex}(\alpha,\xi)=(8-2\xi)\alpha.
\end{align}
The flux per unit cell thus stays at $8\alpha$; $\xi$ only redistributes it, and the homogeneous configuration is recovered at $\xi=1$. The essential requirement is that the optical coupling distinguishes the two $BC$ links within the unit cell. Such spatially selective phases can be generated with holographically structured fields, using a spatial light modulator \cite{mcgloin2003applications,gauthier2021dynamic} or a DMD~\cite{gauthier2016direct,ott2016single,smith2021generation,deng2022maximizing,schroff2023accurate}, together with phase-coherent Raman beams.

\subsection{Effective hopping amplitude from LAT}

The LAT-induced hopping is obtained by projecting the Raman coupling onto the Wannier functions $w_i(\mathbf x)$ of the untilted lattice, centered at the sites $\mathbf X_i$; they are orthonormal and, since the lattice potential is real, can be chosen real. The tilt enters only through the onsite energies $E_i$ introduced above; its distortion of the Wannier functions is of relative order $\mathcal Ea/E_{\rm gap}$, with $E_{\rm gap}$ the gap to the next band, and is neglected.
For a Raman pair with $(\Delta\mathbf k,\Delta\omega)$ the coupling is $V_R=V_0\sin(\Delta\mathbf k\cdot\mathbf x-\Delta\omega t+\Theta)$. With $\mathbf x=\mathbf x'+\mathbf X^c_{ij}$, $\mathbf X^c_{ij}=(\mathbf X_i+\mathbf X_j)/2$, and $\phi_{ij}=\Delta\mathbf k\cdot\mathbf X^c_{ij}+\Theta$, its matrix element between neighboring sites is
\begin{align}
\mathcal K_{ij}(t)&=V_0\,{\rm Im}\Big[\tilde g_{ij}(\Delta\mathbf k)\,e^{i(\phi_{ij}-\Delta\omega t)}\Big], \nonumber\\
\tilde g_{ij}(\mathbf k)&=\int d^2\mathbf x'\,g_{ij}(\mathbf x')\,e^{i\mathbf k\cdot\mathbf x'},
\end{align}
where $g_{ij}(\mathbf x')=w_i(\mathbf x'+\mathbf X^c_{ij})\,w_j(\mathbf x'+\mathbf X^c_{ij})$ is real.

In the interaction picture with respect to onsite energies, $\mathcal K_{ij}$ acquires the factor $e^{i(E_i-E_j)t}$. At the resonance $\Delta\omega=E_i-E_j>0$, one term of $\mathcal K_{ij}$ becomes static, and the remaining, rapidly oscillating term is dropped in the rotating-wave approximation, yielding
\begin{align}\label{eq:Jij}
\mathcal J_{ij}=-\frac{iV_0}{2}\,\tilde g_{ij}(\Delta\mathbf k)\,e^{i\phi_{ij}}, ~ \mathcal J_{ji}=\mathcal J_{ij}^{*}.
\end{align}
The $AC$ bonds, for which $E_i=E_j$, are not driven and keep their real bare hopping. We assume that the diagonal elements of $V_R$ are small, $V_0/\Delta\omega\ll1$, so that the onsite modulation they generate renormalizes the hoppings only at higher order, and that $V_0$ is chosen such that the magnitudes $|\mathcal J_{ij}|$ are comparable to the bare $AC$ hopping.

Writing $\tilde g_{ij}(\Delta\mathbf k)=|\tilde g_{ij}|e^{i\gamma_{ij}}$, the resonant hopping carries the phase $\phi_{ij}+\gamma_{ij}-\pi/2$. The flux through a plaquette is the sum of the phases of its directed hops, with a hop down the tilt counting minus the phase of the reverse hop, so the constant $-\pi/2$ drops out of the triangles and the hexagon. The remaining phases are the Raman contribution $\phi_{ij}$, controlled by $\Delta\mathbf k$ and $\Theta$, and the intrinsic contribution $\gamma_{ij}$, which is set by the Wannier functions and the lattice geometry. 
For a spatially structured Raman phase $\Theta\to\Theta(\mathbf x)$ varying slowly over a bond, $\phi_{ij}=\Delta\mathbf k\cdot\mathbf X^c_{ij}+\Theta(\mathbf X^c_{ij})$, so the bond phase has a contribution from the momentum transfer and an independently controllable contribution from the optical phase profile.
We treat the $\gamma_{ij}$ as fixed, cell-independent offsets and assume that they are compensated by the bond-selective part of $\Theta(\mathbf x)$, so that the net bond phases are $2\pi/\Phi_0$ times those of Eq.~\eqref{bondphase1}, and the resultant flux configuration of the lattice respects Eq.~\eqref{fluxkag}. 

A uniform phase offset of a Raman pair enters the $AB$ and $BC$ bonds of its triangles equally and drops out of the flux, so the scheme requires optical phase control that distinguishes bonds within a unit cell, which is demanding for diffraction-limited addressing. The $\gamma_{ij}$, by contrast, need not be computed from the lattice parameters; they can be calibrated {\it in situ}. Since the $\gamma_{ij}$ are cell-independent, they shift the fluxes of the upper and lower triangles by two constants, while their contributions to the flux per structural unit cell (SUC) cancel, each driven bond being traversed in opposite directions by its triangle and its hexagon. The two optical phases $\theta_1$ and $\theta_2$ can therefore null both offsets, and the SUSY point $\xi=0$ itself provides the calibration signal: flux-free triangles are identified by the collapse of the ALL series in the dynamical structure factor (Sec.~\ref{sec:dsf}) and by the $\Phi_0/8$ periodicity of the spectrum at $\xi=0$ (Sec.~\ref{sec:hr}).

\section{Hofstadter--Rammal spectra in inhomogeneous fluxes}\label{sec:hr}

\begin{figure*}
\centering
    \includegraphics[width=2.07\columnwidth]{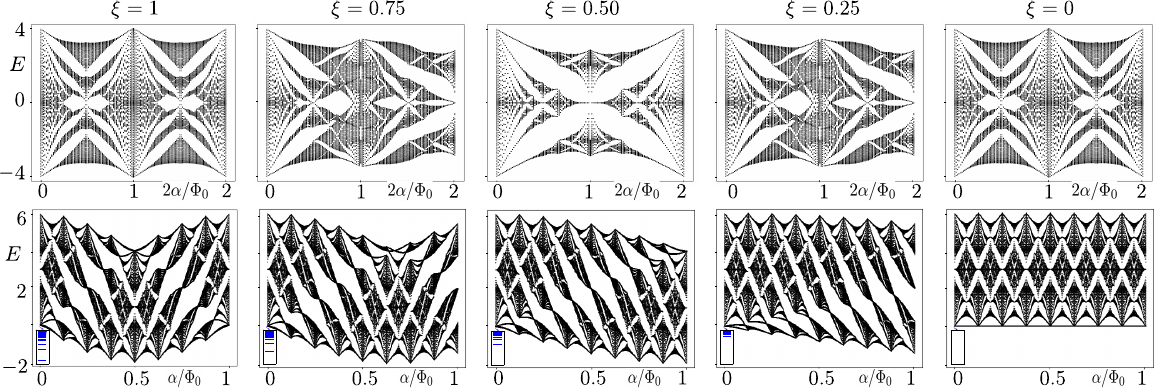}
    \caption{Eigenvalues at the $\Gamma$ point of the MBZ for several values of the flux-modulation parameter $\xi$, for which the homogeneous flux corresponds to $\xi=1/2$ on the square lattice and to $\xi=1$ on the kagome lattice. Upper row: square lattice with a $2q\times2$ magnetic unit cell and $q=137$, as a function of $2\alpha/\Phi_0=p/q$, where $\alpha$ is the mean flux per plaquette; the window $2\alpha/\Phi_0\in(0,2)$ covers $\alpha\in(0,\Phi_0)$. Lower row: kagome lattice with $q=67$, as a function of $\alpha/\Phi_0=p/8q$, where $\alpha$ is the triangle flux at $\xi=1$. For the square lattice, the spectra at $\xi$ and $1-\xi$ coincide at $\Gamma$, since the reflection $m\to-m$ exchanges the two plaquette types. For the kagome lattice, the triangles and the hexagons are inequivalent and cannot be exchanged, so the panels at $\xi$ and $1-\xi$ differ, and there is no chiral symmetry. Inset (kagome): zoom on the ALLs below the flat band at small $\alpha$ set by $p=1$ along with their evolution with $\xi$ (within the energy window $[0,-0.0065]$ covering all ALLs). The degenerate manifold at $\xi=0$ is at $E=0$. }
    \label{fig3}
\end{figure*}

Relative to the homogeneous flux case, the modulated flux alters the HR spectrum differently on a bipartite lattice and on one with triangular plaquettes and a flat band. Fig.~\ref{fig3} shows the eigenvalues at the $\Gamma$ point of the magnetic Brillouin zone (MBZ) for the square lattice (upper row, versus $2\alpha/\Phi_0$) and the kagome lattice (lower row, versus $\alpha/\Phi_0$); both rows cover $\alpha\in(0,\Phi_0)$. Since $\Gamma$ is time-reversal invariant, the symmetries that map it onto itself can be read off directly.

\subsection{Square lattice}

For $2\alpha/\Phi_0=p/q$ with coprimes $p,q$, the $2q$ plaquettes of one period thread a total flux $p\Phi_0$, so $\theta_{m+2q}(\alpha,\xi)=\theta_m(\alpha,\xi)+p\Phi_0$ ($m$ the column index) and the Peierls phases are periodic over $L=2q$ columns. The Landau gauge preserves translation invariance along $y$, so $k_y$ remains a good quantum number.

The square lattice is bipartite, with sublattice operator $(-1)^{m+n}$ ($n$ the row index). This operator is compatible with the even cell length $2q$ along $x$, but for a one-site cell along $y$ it maps $k_y\to k_y+\pi$ and therefore does not act within a single Bloch Hamiltonian. We adopt a $2q\times2$ magnetic unit cell with sites $(m,s)$, $s=0,1$ labeling the two rows, which yields $4q$ bands and makes $\Gamma_{\rm sub}=(-1)^{m+s}$ a genuine chiral operator at every $\mathbf k$. The Peierls phases only decorate the hoppings and generate no onsite terms, so $\Gamma_{\rm sub}H(\mathbf k)\Gamma_{\rm sub}=-H(\mathbf k)$ for all $\mathbf k$, $\xi$ and $\alpha$, and the spectrum is symmetric about $E=0$ (upper row of Fig.~\ref{fig3}).

The dependence on $\alpha$ follows from how the column phases $\theta_m(\alpha,\xi)$ of Eq.~\eqref{dimerized_phase} transform:
\begin{align}
\theta_m(\Phi_0-\alpha,\xi)\equiv{}&-\theta_m(\alpha,\xi)+\Phi_0(2\xi-1)s_m \pmod{\Phi_0},
\end{align}
that is, the time-reversed configuration plus an extra phase $\Phi_0(2\xi-1)$ on every even column, which is trivial if and only if $2\xi\in\mathbb Z$. For $\xi=0,0.5,1$ the tight-binding Hamiltonian is therefore invariant under $\alpha\to\alpha+\Phi_0$. Time reversal maps $\alpha\to-\alpha$ (mod $\Phi_0$) and $\mathbf k\to-\mathbf k$, which leaves $\Gamma$ invariant, so the spectrum is mirror symmetric about $\alpha=\Phi_0/2$, the usual HR reflection $\phi\to1-\phi$ with $\phi=\alpha/\Phi_0$. 

For $\xi=0$ and $1$ there is more structure. Here $\theta_m=\alpha\,n_m$ with $n_m=m\mp s_m$, an odd integer for every $m$, so $\alpha\to\alpha+\Phi_0/2$ shifts every vertical-bond phase by $n_m\Phi_0/2\equiv\Phi_0/2$ and flips the sign of all vertical hoppings. This can be compensated by the gauge transformation $|m,s{=}1\rangle\to-|m,s{=}1\rangle$ on the second row of the magnetic unit cell, which flips all vertical bonds back and leaves the horizontal ones unchanged. The spectrum then has period $\Phi_0/2$ in $\alpha$ and is also mirror symmetric about $\alpha=\Phi_0/4$ (at $2\alpha/\Phi_0=0.5$ and $1.5$). At $\xi=0.5$, by contrast, $\theta_m=\alpha m$, so the shift adds $\Phi_0/2$ only on odd columns; this changes the flux $\alpha$ of every plaquette by $\Phi_0/2$ and cannot be undone by a gauge transformation. In all three cases, the mirror symmetry about $\Phi_0/2$ makes the window $\alpha\in[\Phi_0/2,\Phi_0]$ redundant; for $\xi=0$ and $1$, the additional period $\Phi_0/2$ and mirror about $\Phi_0/4$ reduce it further to $[0,\Phi_0/4]$.

For generic $\xi$ (i.e., $\xi\neq0,1/2,1$; for example $\xi=0.25,0.75$), the extra phase $\Phi_0(2\xi-1)$ is nontrivial, and the spectra in the upper row of Fig.~\ref{fig3} are indeed not mirror symmetric about $\alpha=\Phi_0/2$, the center of each panel. For the same reason, $\alpha\to\alpha+\Phi_0$ shifts the plaquette fluxes $(2\alpha(1-\xi),2\alpha\xi)$ by $(2(1-\xi)\Phi_0,2\xi\Phi_0)$, which are not integer multiples of $\Phi_0$ for $2\xi\notin\mathbb Z$, so the Hamiltonian is not $\Phi_0$-periodic either. For $\xi=0.25$ and $0.75$, however, $\alpha\to\alpha+2\Phi_0$ shifts the fluxes by $(4(1-\xi)\Phi_0,4\xi\Phi_0)$, which are integer multiples of $\Phi_0$, so the Hamiltonian has period $2\Phi_0$ (the spectral period can only be shorter) and, by time reversal, the spectrum is mirror symmetric about $\alpha=\Phi_0$. This mirror line lies at the right edge of the plotted window; hence no mirror symmetry is visible within the panels of Fig.~\ref{fig3}, and the window $\alpha\in(0,\Phi_0)$ already contains all independent content.

Finally, the configurations at $\xi$ and $1-\xi$ are related by the reflection $m\to-m$ about a lattice site combined with time reversal: $\theta_{-m}(\xi)=-\theta_m(1-\xi)$, so that $\Phi^{\rm pl}_{-m-1}(1-\xi)=\Phi^{\rm pl}_m(\xi)$. Since $\Phi^{\rm pl}_m$ depends on $m$ only through its parity, this implies $\Phi^{\rm pl}_m(\xi)=\Phi^{\rm pl}_{m+1}(1-\xi)$, i.e., $\xi\to1-\xi$ exchanges even and odd plaquettes. Both operations map $\Gamma$ onto itself, so the spectra at $\xi$ and $1-\xi$ coincide at every time-reversal-invariant momentum, in particular at $\Gamma$. In Fig.~\ref{fig3}, the panels at $\xi=0$ and $1$ coincide, as do those at $\xi=0.25$ and $0.75$, although the latter show no mirror symmetry within the panels.

In summary, the $\Gamma$-point spectra show three symmetry features: the chiral symmetry $E\to-E$, present for all $\xi$ and $\alpha$; the $\xi\leftrightarrow1-\xi$ equivalence; and the periodicity and reflection in $\alpha$, which depend on $\xi$ and are most constrained at $\xi=0,1/2,1$.

\begin{figure*}
 \centering
 \includegraphics[width=2.07\columnwidth]{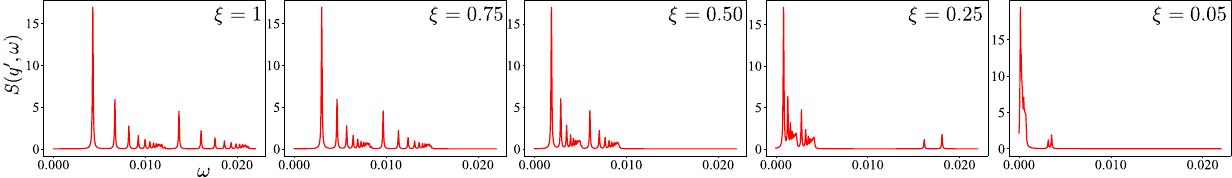}
 \caption{$S(\mathbf Q,\omega)$ at $T=0$ probing ALLs for $q=17$, $p=1$, $\mathbf Q=\mathbf b_2/(qN_2)$, the smallest nonzero momentum of the $\mathbf k$-mesh along the reciprocal-lattice vector $\mathbf b_2$ ($N_1=150$, $N_2=50$), and $\eta=5\times10^{-5}t$, with the two lowest bands filled, for various values of the flux modulation parameter $\xi$. The peaks form the two families of Eq.~\eqref{eq:series}, which are compressed toward lower frequency as $\xi$ decreases; at $\xi=0.25$, the weaker peaks near $\omega\simeq0.016$--$0.018\,t$ are transitions into the lowest dispersive LL. Frequencies are in units of $t$.}
 \label{fig:dsf_panels}
\end{figure*}

\subsection{Kagome lattice}

We now impose the same redistribution of flux on the kagome lattice, which differs from the square lattice in two relevant ways: the triangular plaquettes make it non-bipartite, and its zero-field band structure hosts a singular flat band. With the nn Hamiltonian $H_{\rm k}=t\sum_{\langle ij\rangle}\left(e^{i\phi_{ij}}c_i^\dagger c_j+{\rm h.c.}\right)+\mu\sum_ic_i^\dagger c_i$, and $t>0$, the zero-field flat band lies at the bottom of the spectrum and touches a quadratic band at $\Gamma$. The chemical potential $\mu=2t$ places it at $E=0$, with the dispersive bands in $(0,6t]$ at zero field. Following Ref.~\cite{Datta2025}, the two triangles of the SUC carry flux $\Phi_\triangle=\xi\alpha$ each, and the hexagon carries $\Phi_{\rm hex}=(8-2\xi)\alpha$, so $\Phi_{\rm SUC}=8\alpha$ is independent of $\xi$. Here $\alpha$ is the triangle flux at $\xi=1$, the homogeneous case.

The spectra in Fig.~\ref{fig3} are computed in the gauge of Ref.~\cite{Datta2025}, in which the Peierls phases depend on the position of the SUC only along $\mathbf a_1=2a(1,0)$, so the magnetic unit cell is a $q$-fold enlarged along $\mathbf a_1$ while ${\mathbf k}\cdot{\mathbf a_2}$ with $\mathbf a_2=2a(1/2,\sqrt3/2)$ is a good quantum number. For the $l$-th SUC ($l=1,\dots,q$) the nonzero bond phases are $2\pi\varphi^{(i)}_l/\Phi_0$, with $\varphi^{(1)}_l=8l\alpha$, $\varphi^{(2)}_l=\xi\alpha$ and $\varphi^{(3)}_l=[\xi-8(l+1)]\alpha$, yielding the desired flux configuration $\Phi_\triangle=\xi\alpha$ and $\Phi_{\rm hex}=(8-2\xi)\alpha$ independently of $l$. For $8\alpha/\Phi_0=p/q$, the magnetic cell contains $q$ SUCs (each threading $8\alpha$ flux), and the lower row of Fig.~\ref{fig3} shows the $3q$ levels at $\mathbf k=0$ for $p=1,\dots,8q-1$ and several $\xi\in[0,1]$.

The kagome lattice is not bipartite: its triangles are odd cycles, so no sublattice symmetry exists, and the HR spectrum has a symmetry structure distinct from that of the square lattice. Moreover, $\xi$ transfers flux between triangles and hexagons, which no lattice symmetry interchanges; the $\xi\leftrightarrow1-\xi$ equivalence of the square lattice therefore has no counterpart, and the panels at $\xi$ and $1-\xi$ in Fig.~\ref{fig3} do not coincide. The $\Gamma$-point spectrum obeys $E(\alpha)=E(-\alpha)$. The Peierls phases are invariant under $\alpha\to\alpha+\delta$ if and only if $\xi\delta$ and $8\delta$ are integer multiples of $\Phi_0$, which fixes the periodicity. At $\xi=1$ this gives period $\Phi_0$ and the HR mirror about $\alpha=\Phi_0/2$. At $\xi=0$ the triangles are flux-free and the period shrinks to $\Phi_0/8$, so that $(0,\Phi_0)$ contains eight copies of the same pattern, each mirror symmetric about its center. For rational $\xi=r/s$ the period is at most $s\Phi_0$ (for example $2\Phi_0$ at $\xi=1/2$ and $4\Phi_0$ at $\xi=1/4,3/4$), with the mirror at half the period. The mirror about $\alpha=\Phi_0/2$ is guaranteed only for $\xi\in\mathbb Z$; for $0<\xi<1$ it is not observed, and the window $(0,\Phi_0)$ shows only a fraction of the period, so the panels at intermediate $\xi$ are visibly tilted.

\noindent
{\bf Anomalous Landau levels.} The effect of the modulation on the ALLs is the central result of this section. At $\xi=1$, the inset of Fig.~\ref{fig3} shows the levels emerging below the flat band ($E=0$) at small $\alpha$. They fan out from $\alpha=0$ with slopes that decrease with the level index $n$. 
In the continuum limit, their energies are known exactly~\cite{Datta2025},
\begin{align}
E_n=-E_0\Big[\sqrt{n(n+1)+\tfrac{(1+\xi)^2}{4}}-n-\tfrac12\Big],\quad n\geq1,
\label{eq:allscale}
\end{align}
with $E_0=ta^2/l^2=(8\pi/\sqrt3)(\alpha/\Phi_0)\,t$, where $l$ is the magnetic length, $1/l^2=8\pi(\alpha/\Phi_0)/(\sqrt3a^2)$, and $n$ counting the levels in order of energy; for $n\gg1$, $E_n\simeq-E_0\,\xi(2+\xi)/(8n)$, unveiling an anomalous quantization of LLs associated with the singular flat band~\cite{Rhim2020, Datta2025}. We denote the lowest level by $\epsilon_0\equiv E_1$, which sets the width of the tower. This energy scale is tiny compared with the bandwidth and is resolved in the inset (in the lower row of Fig.~\ref{fig3}). As $\xi$ is lowered at fixed $\alpha$, $|\epsilon_0|$ decreases nearly linearly, and the whole tower is compressed toward the flat-band energy: at large $n$, the levels scale by $\xi(2+\xi)/3$, through an effective quantum distance $d_{\rm eff}=d_{\rm max}\xi(2+\xi)/3$~\cite{Datta2025}, where $d_{\rm max}=1$ is the maximal quantum distance of the kagome flat band~\cite{Rhim2020}.

The squeezing is controlled by the triangle flux: $\xi\alpha$ measures how far the lowest band departs from exact flatness, while the hexagon flux $(8-2\xi)\alpha$ compensates it at fixed flux per SUC, $\Phi_{\rm SUC}$.
At $\xi=0$, the tower collapses onto $E=0$, regardless of $\alpha$. In the lower right panel of Fig.~\ref{fig3}, this appears as a flat line at the lower edge of the spectrum, while the rest of the pattern recurs eight times in $(0,\Phi_0)$. The line is the $q$-fold zero-mode manifold of $H_{\rm k}$, which is positive semidefinite at $\xi=0$ with $\mu=2t$ and has one zero mode per SUC; its origin is an emergent lattice supersymmetry \cite{Datta2025}, which we do not rederive here. The flatness is consistent with compact localization of the lowest band: for flux-free triangles the destructive interference around each hexagon is not spoiled by the field, whereas for $\xi>0$ a leakage amplitude, proportional to $\xi\alpha$ for small $\xi\alpha$, spoils the localization and spreads the zero-energy manifold into the ALL tower below the flat band~\cite{Datta2025}. The collapse is the lattice analog of an Aharonov--Casher manifold, reminiscent of Aharonov--Bohm caging but sustained by an inhomogeneous flux and confined to the lowest band. Since $\epsilon_0$ sets the energy of the excitations between ALLs, its evolution with $\xi$ determines the response studied in Sec.~\ref{sec:dsf}.

\section{Spectroscopic detection of the Landau levels}\label{sec:dsf}

In optical lattices, where the synthetic field is imposed by LAT, LLs can be inferred from observables. Besides probes of the density profile and of transport, spectroscopic probes of the excitation spectrum, such as Bragg and lattice-modulation spectroscopy~\cite{Stenger1999, ernst2010probing, stoferle2004transition}, access the LL structure through the energies of transitions between the MBBs. For conventional LLs, such as $E_n\propto n$, these transitions are equally spaced, and the response is dominated by a single cyclotron peak. For the ALLs of Eq.~\eqref{eq:allscale}, they instead form a nonequidistant series, so the response carries the fingerprint of the anomalous quantization and of its modulation by $\xi$. A density modulation of wave vector $\mathbf Q$ couples to the density operator and measures the DSF $S(\mathbf Q,\omega)$, whose peaks lie at the energies of density-induced transitions between the MBBs.

For noninteracting spinless fermions in $H_{\rm k}$, we label the $3q$ MBBs $E_{\nu\mathbf k}$ in ascending energy, so that $\nu=1$ is the lowest ALL. With the lowest $N_{\rm occ}$ bands filled, the DSF at $T=0$ and $\omega>0$ is
\begin{align}
S(\mathbf Q,\omega)={}&\frac{1}{N_{\mathbf k}}\sum_{\mathbf k}\sum_{\nu\le N_{\rm occ}<\nu'}\left|M_{\nu'\nu}\right|^{2} \delta\!\left(\omega-\Delta E_{\nu'\nu}\right),
\label{eq:DSF_T0}
\end{align}
where $N_{\mathbf k}=N_1N_2$ is the number of $\mathbf k$ points of a uniform $N_1\times N_2$ mesh along the reciprocal-lattice vectors $\mathbf b_1$ and $\mathbf b_2$ of the MBZ, $\Delta E_{\nu'\nu}=E_{\nu',\mathbf k+\mathbf Q}-E_{\nu\mathbf k}$ with $\mathbf Q$ commensurate with the mesh, and the weight of each transition is set by the matrix element of the density operator,
\begin{align}
M_{\nu'\nu}(\mathbf k,\mathbf Q)&=\langle u_{\nu',\mathbf k+\mathbf Q}|\,\rho_{\mathbf Q}\,|u_{\nu,\mathbf k}\rangle,~\rho_{\mathbf Q}=\sum_{s=1}^{3q}e^{i\mathbf Q\cdot\mathbf r_s}\,|s\rangle\langle s|.
\label{eq:Mmn}
\end{align}
Here $|u_{\nu\mathbf k}\rangle$ are the eigenvectors of the kagome Bloch Hamiltonian $H(\mathbf k)$ with energies $E_{\nu\mathbf k}$, and $\mathbf r_s$ are the site positions of the $3q$-site cell. Numerically, the delta function is replaced by a Lorentzian of width $\eta$. The energies $\Delta E_{\nu'\nu}$ are set by the band dispersion and the weights $|M_{\nu'\nu}|^2$ by the Bloch eigenstates.

Since $S$ probes energy \emph{differences}, the anomalous quantization appears as a Rydberg-like series. The bands being nearly flat, $\Delta E_{\nu'\nu}$ depends weakly on $\mathbf k$, and the transition $\nu\to\nu'$ appears at
\begin{align}
\omega_{\nu\nu'}&\simeq E_{\nu'}-E_\nu,
\label{eq:series}
\end{align}
with $E_n$ from Eq.~\eqref{eq:allscale}. For fixed $\nu$, the peaks become denser with increasing $\nu'$ and converge to $|E_\nu|$, the energy needed to excite a particle from level $\nu$ to the flat-band energy $E=0$. For $N_{\rm occ}=2$ and $\xi=1$, the family from $\nu=2$ spans $0.040\,E_0$ to $0.146\,E_0$ and that from $\nu=1$ spans $0.127\,E_0$ to $0.232\,E_0$; the lowest peak, $(E_{N_{\rm occ}+1}-E_{N_{\rm occ}})$, is the gap between the last occupied and the first empty level and thus sets the frequency resolution and temperature required to resolve the series, as discussed below. A weaker branch arises from transitions into the lowest dispersive LL, which lies at $\xi E_0/2$~\cite{Datta2025}, at $\omega\simeq\xi E_0/2-E_\nu$.

\begin{figure}
 \centering
 \includegraphics[width=\columnwidth]{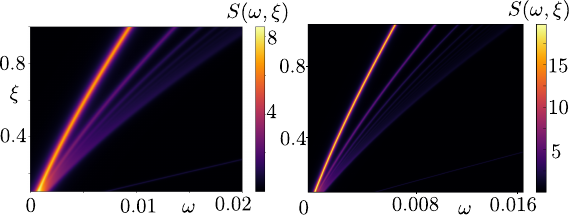}
 \caption{$S(\omega,\xi)$ for $q=17$ (left, $\eta=3\times10^{-4}t$) and $q=23$ (right, $\eta=10^{-4}t$), with the two lowest bands filled and $\mathbf Q=\mathbf b_2/(qN_2)$ as in Fig.~\ref{fig:dsf_panels}. The ridges are the peak series of Fig.~\ref{fig:dsf_panels} and converge to $\omega\to0$ as $\xi\to0$; the faint line at the lower right is the branch of transitions into the lowest dispersive LL, which vanishes linearly as $\xi\to0$.}
 \label{fig:dsf_map}
\end{figure}

Figure~\ref{fig:dsf_panels} shows $S(\mathbf Q,\omega)$ for $q=17$ and $N_{\rm occ}=2$ out of $3q$ bands. The peaks form two families converging to two series limits, the lower one arising from $\nu=2$ and the upper from $\nu=1$. As $\xi$ decreases, the whole pattern is compressed toward lower frequency, with the relative peak heights nearly unchanged, so the modulation rescales the transition energies and hardly redistributes the weights $|M_{\nu'\nu}|^2$. Figure~\ref{fig:dsf_map} shows the evolution versus $\xi$ for $q=17$ and $q=23$: the ridges converge to $\omega\to0$ as $\xi\to0$, and are shifted to lower frequency for the larger $q$, i.e.\ for the weaker flux. The weak branch, the transitions into the lowest dispersive LL, extrapolates linearly to zero as $\xi\to0$, where that level joins the zero-energy manifold. Resolving the series requires $\eta$ below the lowest transition energy, which decreases with $\xi$, so we restrict to $\xi\ge0.05$. At $\xi=0$, the occupied levels merge into the degenerate zero-mode manifold, and the response depends on how it is filled.

In an optical-lattice realization, all ALL frequencies scale with $E_0=(8\pi/\sqrt3)(\alpha/\Phi_0)\,t\simeq14.5\,(\alpha/\Phi_0)\,t$ and are much smaller than $t$. For $q=17$ ($\alpha/\Phi_0=1/136$) and $\xi=1$, the lowest peak of Fig.~\ref{fig:dsf_panels} lies at $\omega\approx4.3\times10^{-3}\,t$, i.e., $4.3$\,Hz for $t/h=1$\,kHz; at $\alpha/\Phi_0=1/80$, the lowest ALL lies $\approx42$\,Hz below the flat band and, with one ALL filled, the first transition is at $\approx16$\,Hz. 
Resolving the series requires a frequency resolution, and hence a spectroscopy time, and an effective temperature below the gap $(E_{N_{\rm occ}+1}-E_{N_{\rm occ}})$, here corresponding to sub-nanokelvin temperatures and probe durations of order $0.1$--$1$\,s. These requirements are demanding for driven lattices, where heating limits the coherence time, and they motivate the two routes discussed next.
The first route is to increase the flux: all ALL frequencies grow linearly with $\alpha$, which relaxes both requirements, at the price of fewer ALLs below the flat band, since their number, $q-p$, decreases with $\alpha$; a stronger flux thus resolves a shorter tower at higher frequencies. The second route targets the SUSY point, whose signatures do not rely on the small ALL scale and persist at large flux, where all energy scales are a sizable fraction of $t$: the exact zero-energy manifold, which does not move with $\alpha$, and the $\Phi_0/8$ periodicity of the spectrum at $\xi=0$. These signatures are therefore the natural first targets of an experiment.

\section{Conclusion and outlook}\label{sec:concl}

We have shown how LAT imprints spatially inhomogeneous synthetic fluxes on optical lattices, through two coherent Raman processes on the square lattice and bond-selective Raman phases on the kagome lattice. The two lattices respond differently to the flux modulation (dictated by the parameter $\xi$): the bipartite square lattice keeps its chiral symmetry and the $\xi\leftrightarrow1-\xi$ equivalence in its HR spectrum, whereas on the kagome lattice the redistribution compresses the ALLs of the singular flat band and collapses them onto a zero-energy manifold when the triangles become flux-free. In the DSF, accessible by Bragg spectroscopy, this evolution appears as a Rydberg-like series of peaks that is compressed toward zero frequency as $\xi$ decreases.

These results identify the intracell distribution of flux as a control parameter in its own right, beyond the mean flux that defines the Hofstadter problem. On the kagome lattice, it tunes the ALL bandwidth continuously to zero at fixed band structure, {\it in situ} and reversibly. This makes the regime in which interactions dominate the kinetic energy of the ALLs directly accessible; the role of interactions in partially filled ALLs, where the nearly degenerate manifold may favor correlated states, is a natural next question. The SUSY point, set by flux-free triangles, is an experimental sweet spot: its signatures persist at large flux, and, since they appear only when both triangle fluxes vanish, they also calibrate the optical phases that set $\xi$ (Sec.~\ref{sec:kaglat}).

The construction is not specific to the kagome lattice. Any lattice whose Hamiltonian factorizes over a root graph with closed loops \cite{roychowdhury2024supersymmetry} admits flux patterns that leave those loops flux-free, and the same mechanism of collapse can be explored there. Beyond spectroscopy, the topological character of the ALLs could be probed through transport or Hall drift measurements. Our analysis is limited to noninteracting fermions at $T=0$ and neglects Floquet heating, trap inhomogeneity, and finite size; including these effects is needed for a quantitative comparison with experiment.

\begin{acknowledgments}
KR acknowledges support from the Prime Minister Early Career Research Grant (PMECRG)-funded project (File No. ANRF/ECRG/2024/006260/PMS). 
\end{acknowledgments}

\bibliographystyle{apsrev4-2}
\bibliography{reference}

\end{document}